\documentclass[
]{ceurart}

\usepackage{listings}
\usepackage{float} 
\usepackage{listings}

\usepackage{placeins}

\usepackage{tikz}

\newcommand{\bcircled}[1]{\tikz[baseline=(char.base)]{
		\node[shape=circle,fill=red,draw=red,inner sep=1.5pt] (char)
		{\textcolor{white}{\scriptsize\textbf{#1}}};}}

\usepackage{graphicx}
\usepackage{float}      
\usepackage{placeins}   
\usepackage{caption}
\usepackage{subcaption}

\usepackage{afterpage}  

\begin{document}

\copyrightyear{2022}
\copyrightclause{Copyright for this paper by its authors.
  Use permitted under Creative Commons License Attribution 4.0
  International (CC BY 4.0).}

\conference{Submitted to Posters \& Tools Track, Co-located with REFSQ 2026.}

\title{RITA: A Tool for Automated Requirements Classification and Specification from Online User Feedback}


\author[1]{Manjeshwar Aniruddh Mallya}[%
email=mallyaaniruddh@gmail.com,
]

\author[2]{Alessio Ferrari}[%
orcid=0000-0002-0636-5663,
email=alessio.ferrari@ucd.ie,
]

\author[3]{Mohammad Amin Zadenoori}[%
orcid=0000-0003-4591-153X,
email=amin.zadenoori@unipd.it,
]

\author[1]{Jacek Dąbrowski}[%
orcid=0000-0003-3392-0690,
email=jacek.dabrowski@lero.ie,
]
\cormark[1]

\address[1]{Lero, the Research Ireland Centre for Software, University of Limerick, Ireland}

\address[2]{University College Dublin (UCD), Ireland}

\address[3]{University of Padova, Italy}

\cortext[1]{Corresponding author.}

\begin{abstract}
\textbf{\textbf{Context and motivation.}} Online user feedback is a valuable resource for requirements engineering, but its volume and noise make analysis difficult. Existing tools support individual feedback analysis tasks, but their capabilities are rarely integrated into end-to-end support. \textbf{Problem.} The lack of end-to-end integration limits the practical adoption of existing RE tools and makes it difficult to assess their real-world usefulness. \textbf{Solution.} To address this challenge, we present \textsc{RITA}, a tool that integrates lightweight open-source large language models into a unified workflow for feedback-driven RE. \textsc{RITA} supports automated request classification, non-functional requirement identification, and natural-language requirements specification generation from online feedback via a user-friendly interface, and integrates with \textsc{Jira} for seamless transfer of requirements specifications to development tools. \textbf{Results and conclusions.} \textsc{RITA} exploits previously evaluated LLM-based RE techniques to efficiently transform raw user feedback into requirements artefacts, helping bridge the gap between research and practice. A demonstration is available at: \url{https://youtu.be/8meCLpwQWV8}.

\end{abstract}


\maketitle

\section{Introduction}

Online user feedback, such as app store reviews or online discussion forums, contains useful information about what users want, what problems they encounter, and how they perceive software quality~\cite{Dabrowski2022}. This information is useful for requirements engineering (RE) activities (e.g., requirements specification)~\cite{Dabrowski2022a}. However, the large volume of feedback and its informal and noisy nature make it difficult for requirements engineers to analyse it manually and turn it into structured requirements artefacts~\cite{AlSubaihin2021}.

A variety of automated techniques have been proposed to support the analysis of user feedback~\cite{Dabrowski2022}. These include methods for classifying feedback, extracting features, and identifying sentiment or topics~\cite{11190331}. More recently, large language models (LLMs) have attracted attention because they can understand natural language and generate text~\cite{zadenoori2025llmsre}. This makes them suitable for tasks such as identifying user requests and drafting requirements specifications~\cite{DBLP:conf/re/PasqualeRPD25}

In our previous study, we empirically evaluated lightweight open-source LLMs on feedback-driven RE tasks, including user request classification, non-functional requirement identification, and requirements specification generation~\cite{Mallya2026FromOnlineUserFeedback}. We showed that such models can support these tasks with moderate success, while also observing limitations related to consistency, accuracy, and practical use.

Although this prior work provided evidence of what LLMs can do on individual RE tasks, it did not address how these techniques could be combined and used within a single tool that fits practitioners’ workflows~\cite{Dabrowski2022a,zadenoori2025llmsre}. Moreover, existing tools typically focus on isolated analysis functions rather than providing end-to-end support that transforms raw user feedback into artefacts that can be used directly in software development~\cite{Dabrowski2022}.

To address this gap, we developed \textsc{RITA}, a tool that integrates multiple LLM-based analysis tasks into a unified workflow for feedback-driven requirements engineering. \textsc{RITA} supports requirements engineers in analysing online user feedback by automatically classifying user requests, identifying non-functional concerns, and generating natural-language requirements specifications. These capabilities are delivered through a user-friendly graphical interface. In addition, the tool is integrated with \textsc{Jira}, enabling generated requirements artefacts to be transferred directly into widely used development tools.

The goal of \textsc{RITA} is to enable the use of LLM-based techniques in practice during early and ongoing RE activities that rely on continuous user feedback. The tool is intended for practitioners who need lightweight support for analysing feedback and generating draft requirements within existing tools.

The key contribution of this paper is the move from isolated, offline evaluation of LLM4RE techniques to their use in a fully integrated end-to-end tool. Unlike our prior work~\cite{Mallya2026FromOnlineUserFeedback}, which studied individual tasks in isolation, this paper introduces \textsc{RITA}, a working system that combines these tasks into a single interactive workflow and connects them to existing software development tools. By providing a working tool rather than only benchmarks, \textsc{RITA} enables studies of how LLM-based RE techniques are used in practice. A demo~\cite{RITA_Demo_YouTube} and the replication package~\cite{RITA_GitHub} are publicly available.

\section{RITA Tool Description}

We now describe the features of the RITA tool from the perspective of its end users. We then present an overview of the user flow, the architecture of the tool, and details of its implementation.

\subsection{Use Cases}

This section describes RITA from the perspective of its users. RITA supports requirements analysts, product managers, and software engineers in converting large collections of textual user feedback into structured software requirements. Within the system boundary, RITA supports the following use cases:

\begin{itemize}
	
	\item \textbf{UC1 – Upload feedback corpus.}  
	The user uploads a dataset of user feedback (e.g., app reviews, survey responses, or issue reports) to be analysed. RITA supports common data formats (txt, csv and xls) and accepts feedback from multiple sources.
	
	\item \textbf{UC2 – Configure classification scheme.}  
	The user defines how the feedback is analysed by selecting a classification type (e.g., User Request Types or Non-Functional Requirements), choosing a large language model, and specifying a prompt strategy. This allows the analysis to be adapted to different project or research goals.
	
	\item \textbf{UC3 – Classify feedback.}  
	RITA processes the uploaded feedback and assigns each entry to one or more requirement-related categories based on the selected configuration, converting raw text into structured requirement labels.
	
	\item \textbf{UC4 – Inspect classification results.}  
	The user reviews the classification output to see how individual feedback items were interpreted, supporting validation and qualitative analysis.
	
	\item \textbf{UC5 – Generate Software Requirements Specifications (SRS).}  
	RITA generates formalised SRS documents from the classified feedback, consolidating and structuring the extracted requirements for documentation and communication.
	
	\item \textbf{UC6 – Generate user stories.}  
	RITA converts classified feedback into agile-style user stories, allowing the user to move from raw feedback to backlog-ready artefacts.
	
	\item \textbf{UC7 – Export requirements artefacts.}  
	The user exports classification results, SRS documents, and user stories in standard formats (e.g., CSV) for further analysis, reporting, or archiving.
	
	\item \textbf{UC8 – Integrate with Jira.}  
	The user transfers selected user stories into Jira as issues so that the generated requirements can be managed within existing development workflows.
	
\end{itemize}

\subsection{User Flow and Screen Presentation}

\begin{figure}[ht]
    \centering
    \includegraphics[
    width=1\textwidth,
    height=1\textheight,
    keepaspectratio
    ]{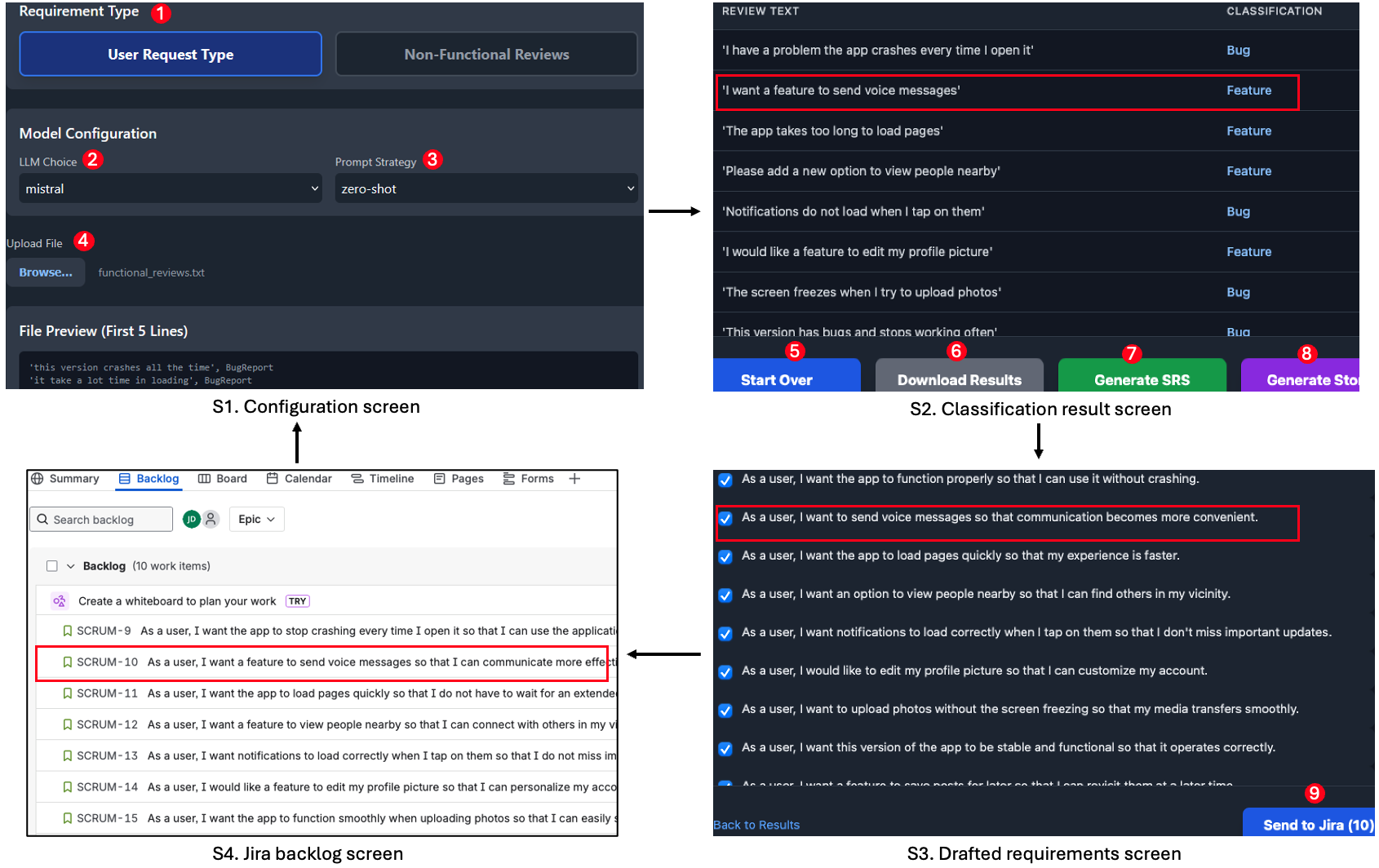} 
    \caption{User Flow and Screen Presentation in RITA Tool and Integrated Jira Project Management Tool.}
    \label{fig:user_flow}
\end{figure}

The RITA tool guides the user through a sequence of screens for configuring the analysis (S1), viewing the classification results (S2), and viewing requirement drafts (S3). The exported output from RITA, in the form of requirements, is then made available in the Jira backlog screen (S4). Figure~\ref{fig:user_flow} presents the user flow and screen presentation, focusing on four main screens. The user flow starts on the configuration screen (see S1). Here, the user selects the classification type \bcircled{1}, the large language model \bcircled{2}, and the prompt strategy \bcircled{3} that control how the feedback is processed. These settings can be changed between runs without re-uploading the data. After configuration, the user uploads the feedback file \bcircled{4}. RITA accepts CSV, Excel, and text files. For tabular inputs, the user selects the column containing the feedback text. Once the file is submitted, the system starts the classification process and displays progress information, such as the number of feedback items processed. The user can cancel the run and restart it with different settings if needed. When processing is complete, the classified feedback is displayed in a structured table on the classification results screen (see S2). Individual entries can be inspected, and the full output can be downloaded. From this screen, users can restart the classification with a different input file or classification type \bcircled{5}, or download the results \bcircled{6}. Users can also proceed to generate higher-level artefacts. RITA can generate Software Requirements Specifications \bcircled{7} and user stories \bcircled{8} from the classified feedback (see S3). After the user stories are generated, the user can select stories of interest and transfer them to Jira \bcircled{9}. These user stories are then created as Jira items and appear in the Jira interface (see S4).

\subsection{Overview of the RITA Architecture}

We now present a high-level overview of the RITA architecture. This section describes the system’s main components, their structural organization, and their deployment in the execution environment. It also explains how these components interact at runtime to realize the intended use cases.

\subsubsection{Component View}

\begin{figure}[!htbp]
    \centering
    \includegraphics[
      width=\textwidth,
      height=0.25\textheight,
      keepaspectratio
    ]{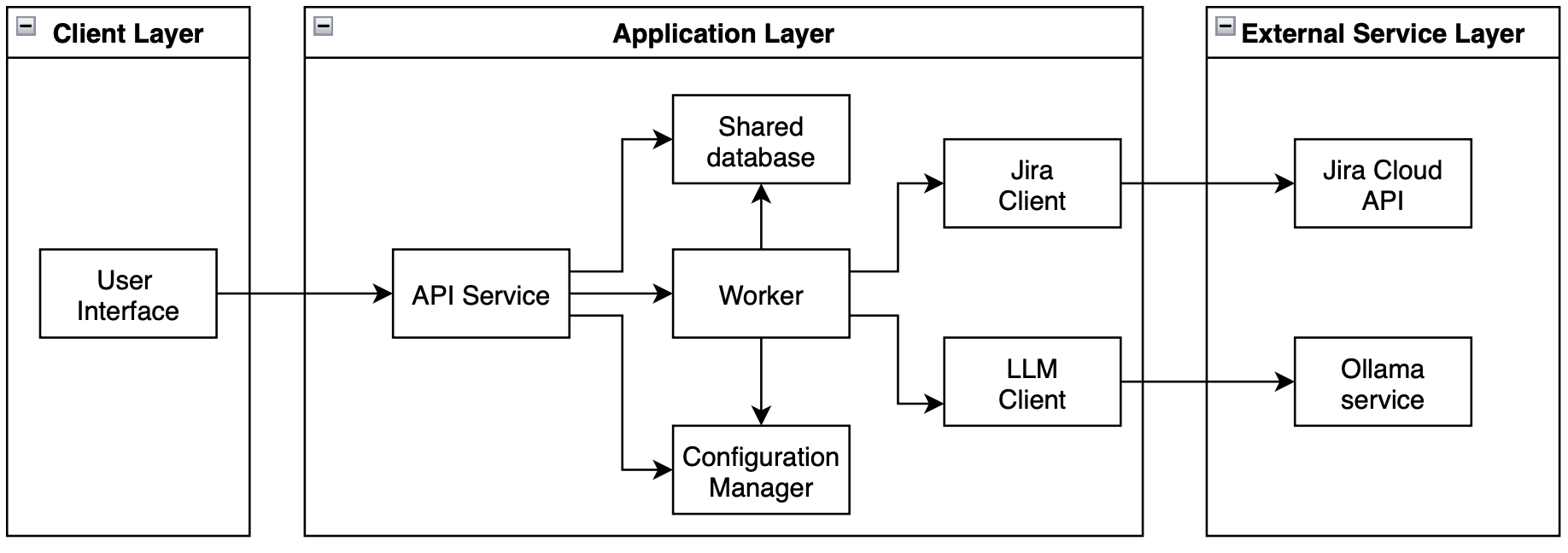}
    \caption{Structural Overview of the RITA Tool. Arrows indicate usage dependencies and point from the invoking component to the component it uses.}
    \label{fig:rita_component_diagram}
\end{figure}

The RITA architecture is organised into three layers: \textbf{client}, \textbf{application}, and \textbf{external service}. It is composed of several components separating user interaction from data analysis and processing. Figure~\ref{fig:rita_component_diagram} presents the RITA component structure and their interconnections; arrows point from components that initiate interactions to the components they use.

The \textbf{client layer} consists of a user interface (UI) that runs in the user’s web browser. This interface provides a web interface for users to upload review files and configure analysis options. It also allows users to initiate classification and requirements generation and to view the generated results. All user interactions are translated into structured HTTP requests, sent to the backend \textit{API service} for processing.

The \textbf{application layer} acts as the central coordination point of the system and consists of six components. The \textit{API service} serves as the entry point to the backend. It is directly connected to the client layer, the \textit{configuration manager}, the \textit{worker service}, and the \textit{shared database}. The API service receives requests from the client layer and validates the inputs. It stores uploaded data, starts background processing, and retrieves results for display in the UI. The \textit{configuration manager} is accessed by both the API service and the worker service. It maintains analysis parameters, model settings, and project-specific options used during execution. The \textit{worker service} is responsible for executing computationally intensive and long-running tasks. These tasks include analysing large collections of review text and generating requirements artefacts. The worker service is invoked by the API service and retrieves the configuration from the configuration manager. During execution, it reads input data from the \textit{shared database} and stores intermediate and final results there. For language model inference, the worker service uses the \textit{LLM client} to communicate with the language model service. It also uses the \textit{JIRA client} to transfer generated requirements artefacts to external Jira systems.
The \textit{shared database} is accessed by both the API service and the worker service. It stores uploaded reviews, intermediate processing states, and generated artefacts. This enables persistent storage and retrieval across user sessions.

The \textbf{external service layer} consists of two components. The \textit{Ollama local service} hosts large language models. It is accessed only through the LLM client to perform automated analysis and text generation. The \textit{Jira Cloud applications} are accessed through the JIRA client. They receive exported user stories and requirements artefacts generated by the RITA tool.

\subsubsection{Use Case Realisation}
\label{sec:use-case-realisation}

\FloatBarrier

\begin{figure}[!htbp]
	\centering
	\includegraphics[
	width=\textwidth,
	keepaspectratio
	]{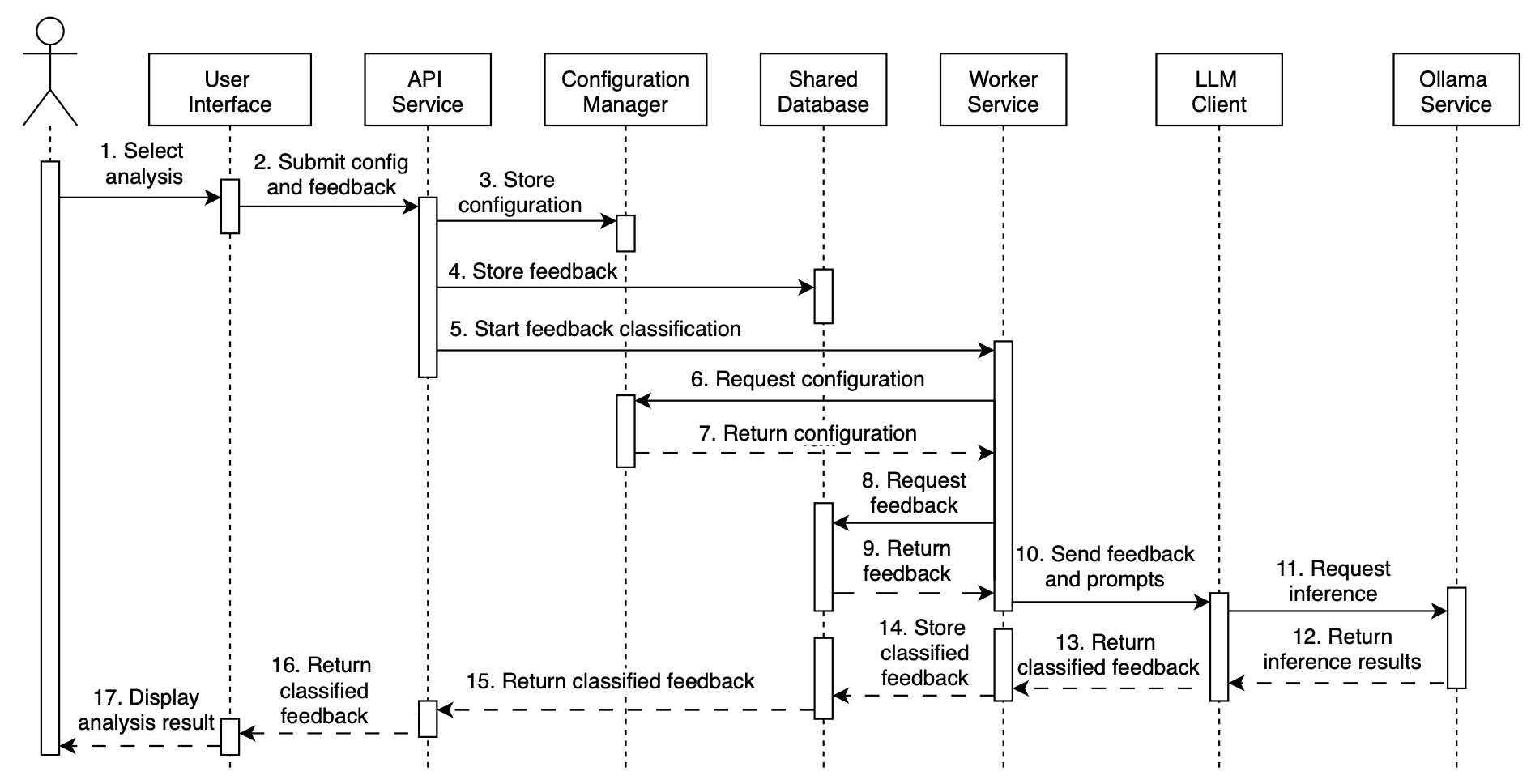}
	\caption{Sequence Diagram of RITA (Review Classification Workflow)}
	\label{fig:rita_sequence_diagram}
\end{figure}

We now describe how RITA realises its two main use cases: feedback classification and requirements specification, and Jira integration and requirements export. The first use case is presented in more detail and is accompanied by a supplementary sequence diagram that illustrates the overall interaction between system components. The second use case is described more briefly, as it follows the same interaction structure.

\paragraph{Feedback Classification and Requirements Specification}

Figure~\ref{fig:rita_sequence_diagram} illustrates the interaction sequence for the feedback classification workflow. The process begins when the user selects the classification configuration and uploads a feedback dataset through the user interface. The user interface forwards the configuration and data to the backend \textit{API service}. The API service stores the configuration via the \textit{configuration manager}, persists the feedback corpus in the \textit{shared database}, and invokes the \textit{worker service} to start the classification job.
The worker service retrieves the active configuration from the configuration manager and requests feedback items from the shared database. Each feedback item is submitted to the language model through the \textit{LLM client}, which forwards the inference request to the \textit{Ollama local service}. The classification results returned by the language model are passed back through the LLM client to the worker service and stored in the shared database. Once available, the API service retrieves the stored classification results from the shared database. The results are then returned or streamed to the user interface, where they are displayed to the user. 

After reviewing the classification results, the user may initiate the generation of higher-level requirements artefacts. This process follows the same interaction pattern as the feedback classification workflow; therefore, a separate sequence diagram is omitted for the sake of brevity. The user interface forwards the generation request to the \textit{API service}, which triggers the \textit{worker service} to retrieve the classified feedback from the \textit{shared database} and the active configuration from the \textit{configuration manager}. The worker service constructs generation prompts and submits them to the language model via the \textit{LLM client}. The generated artefacts, such as SRS and user stories, are stored in the \textit{shared database}. They are later retrieved by the API and displayed in the UI for inspection and export.

\paragraph{Jira Integration and Requirements Export} After reviewing the generated requirements artefacts, the user may choose to export user stories through the user interface. The user interface forwards this request to the backend \textit{API service}, which triggers the \textit{worker service} to initiate the export process. The worker service uses the \textit{JIRA client} to transfer the selected user stories to the \textit{Jira Cloud applications}. The exported user stories are created as Jira issues and become part of the target project’s backlog.

\subsubsection{Deployment View}

RITA is deployed as a self-contained system on a single host machine and is accessed through a web browser. 
The \textbf{client layer} runs in the browser, while the \textbf{application layer} is deployed on the host machine. The \textbf{application layer} includes locally deployed components, including the \textit{LLM client} and the \textit{JIRA client}. These components run on the host machine and act as connectors to external services. Language model inference is performed by the \textit{Ollama local service} in the \textbf{external service layer}. It is accessed through the \textit{LLM client} and runs on the same host machine, ensuring that review data remains local. The \textit{JIRA client} connects from the host machine to the \textit{Jira Cloud applications} in the \textbf{external service layer} over the network. This connection is used to export generated requirements and artefacts into selected Jira projects. These deployment relationships are illustrated in the deployment diagram.

\subsection{Implementation}

RITA is implemented as a lightweight, Docker-containerised system designed to support transparent and reproducible experimentation. The user interface is built with React and provides a stateless workflow for uploading review datasets, configuring classification parameters, and inspecting generated results~\cite{React_UI}. The FastAPI service forms the Python-based backend of the system and implements the core orchestration logic~\cite{FastAPI_Framework}. It exposes REST endpoints for submitting analysis requests and retrieving results. It also provides streaming endpoints for delivering incremental processing output to the user interface. The backend coordinates all execution by invoking a background processing component and managing data flow between system components. Computationally intensive tasks, including feedback classification and requirements generation, are executed by this background processing component. A local SQLite database is used to store uploaded data, intermediate results, and generated artefacts during execution~\cite{SQLite_DB}. Language model inference is performed by a locally hosted Ollama service~\cite{Ollama_Python}. This service is accessed through a dedicated LLM client. The client encapsulates model-specific communication and enables interchangeable inference configurations. Optional integration with Jira Cloud is supported through a JIRA client that exports selected user stories as Jira issues~\cite{Jira_Software}.

\section{Challenges, Next Steps and Evaluation}

\noindent\textbf{Challenges.} RITA is still under development and some functions are not yet stable. The tool generates user stories or SRS that sometimes fail to meet the required structure or level of detail. We will address this by improving prompt design, strengthening output post-processing, and systematically debugging the implementation. The lightweight open-source LLMs currently used limit the accuracy and consistency of the results. These models may misclassify feedback or produce unclear requirements. We plan to improve this by customising models, engineering better prompts, and fine tuning models on domain specific feedback and requirements data. The tool also faces challenges when integrating its outputs into real development workflows. Although RITA supports Jira, the generated artefacts do not always match industrial documentation and backlog standards. We aim to address this by refining export formats and improving the tool based on practitioner feedback.

\noindent\textbf{Next Steps and Evaluation.} The next phase of this work focuses on improving the stability of RITA and the quality of its outputs. We will fix known implementation issues, improve the user interface, and refine the formatting of SRS and user stories. We will also experiment with additional open-source models, improve prompt design, and fine tune or retrain models using requirements and user feedback data. In addition, we will integrate commercial LLMs to increase output accuracy and consistency. We plan to evaluate RITA with requirements engineers, product managers, and software developers using both quantitative and qualitative methods. Quantitatively, we will compare the time required to analyse the same sets of user feedback with and without RITA in order to measure its effect on efficiency. We will also measure the number of requirements identified and the effort required to revise the generated artefacts. Qualitatively, we will collect feedback through interviews and questionnaires to assess usability, perceived usefulness, and confidence in the generated results. These results will allow us to determine whether RITA provides practical value for feedback driven requirements engineering.

\section{Conclusion}

This paper presented \textsc{RITA}, a tool that integrates lightweight open-source LLMs into a unified workflow for feedback-driven RE. Building on our prior empirical work~\cite{Mallya2026FromOnlineUserFeedback}, \textsc{RITA} applies previously evaluated LLMs to analyse online user feedback. It supports user request classification, non-functional requirement identification, and requirements specification generation using a graphical interface; and integration with the \textsc{Jira} project management tool. The main contribution of \textsc{RITA} is its end-to-end support for bridging data-driven RE research with practical RE workflows. By combining multiple feedback analysis tasks in a single tool and transferring generated artefacts directly to \textsc{Jira}, \textsc{RITA} may enable the systematic and structured use of LLM-based techniques in industrial settings. This paper focuses on the tool’s design and capabilities. Future work will evaluate its real-world usefulness and effectiveness. It will also extend the tool with additional feedback analysis features and deeper process integration.

\begin{acknowledgments}
 A major part of this work was conducted as part of the MSc thesis of M. A. Mallya, supervised by J. Dąbrowski~\cite{Mallya2025}. The results contribute to the Prompt Me project~\cite{Dabrowski2024}. This publication has emanated from research jointly funded by Taighde Éireann – Research Ireland under Grant Number 13/RC/2094\_2, and co-funded by the European Union under the Systems, Methods, Context (SyMeCo) programme Grant Agreement Number 101081459. Views and opinions expressed are however those of the author(s) only and do not necessarily reflect those of the European Union or the European Research Executive Agency. Neither the European Union nor the granting authority can 
be held responsible for them.

\end{acknowledgments}


\bibliography{bibliography-refsq-tool}

\appendix

\end{document}